\documentclass[a4paper,11pt]{article}
\usepackage{pos}

\def\cZ{ \mathcal{Z} } 
\def\cS{ \mathcal{S} }

\title{String theory mathematics and matrix data analysis. }
\author*[a]{Sanjaye Ramgoolam }
\affiliation[a]{Queen Mary University of London,\\
  327 Mile End Road, London, United Kingdom }

\emailAdd{s.ramgoolam@qmul.ac.uk}
\abstract{Inspired by matrix techniques in quantum field theory and string theory, we review Permutation Invariant Gaussian Matrix Models (PIGMM), which replace the continuous symmetries of traditional Random Matrix Theory, for $ N \times N$ matrices, with finite permutation symmetry, $S_N$. This symmetry-driven approach reduces  highly multivariate $N^2$-variable matrix data analysis  problems to a rich but tractable space of parameters. The  representation theory of $S_N$ brings a highly correlated quadratic matrix action to a near-diagonal form with $13$ parameters. The invariant observables are parameterised by graphs and their expectation values are computed with Wick contractions implemented algorithmically. We review the successful application of PIGMM for data reduction and anomaly detection in computational linguistics, statistical finance and neural network weights. We conclude with a brief discussion of potential future applications to matrix data analysis tasks that exploit  hadronisation algorithms and the modular structure of collider-physics data.}

\FullConference{23rd International Workshop on Advanced Computing and Analysis Techniques in Physics Research (ACAT2025)\\
8–12 September 2025\\
Hamburg, Germany\\}

\begin{document}
\maketitle

\section{Introduction}

Traditional Random Matrix Theory (TRMT), originating in the works of Wigner, Dyson and Mehta \cite{Wigner1955,Dyson1962,Mehta1967,Wigner1967}, has revealed universal eigenvalue statistics across diverse applications. Gauge-string duality and the AdS/CFT correspondence \cite{Malda,GKP,Witten} motivate TRMT matrix distributions and their multi-matrix generalisations \cite{CJR2001,SR2016}, where the classification of gauge-invariant observables and the computation of their correlators is central.

Permutation Invariant Gaussian Matrix Models (PIGMM) \cite{LMT-2017,PIGMM-2018} draw on both perspectives, replacing continuous matrix symmetry by finite permutation symmetry. They provide symmetry-based data reduction and have been applied to matrix data in computational linguistics, finance and neural-network weights.

\section{From Continuous to Finite Symmetries: TRMT to PIGMM}
\label{sec:trmt_vs_pigmm}

Wigner, Dyson and Mehta showed that regularities in nuclear energy levels could be modelled by simple matrix probability distributions and this has been supported by more recent studies \cite{Wigner1955,Dyson1962,Mehta1967,Wigner1967,Haq1982}. For real symmetric matrices $M=M^{\top}\in\mathbb{R}^{N\times N}$, the Gaussian Orthogonal Ensemble (GOE) is governed by
\begin{equation}
P(M)dM = \mathcal{C}_{N}\exp\left(-\frac{N}{4}\text{Tr}M^{2}\right)\prod_{i\le j}dM_{ij} \, . 
\label{eq:goe_action}
\end{equation}
Here $\mathcal{C}_{N}$ is a normalisation constant, and the measure is invariant under $M\rightarrow AMA^T$ for $A\in O(N)$. TRMT also describes characteristics of chaotic quantum dynamics and financial correlation matrices \cite{Guhr1998,Akemann2011}. Continuous symmetry reduces the otherwise highly multivariate problem to an uncoupled measure, such as Eq.~\eqref{eq:goe_action}, with independent matrix variables.

Many real-world matrix ensembles are defined on a fixed list of discrete labelled entities, while the information of interest is independent of their ordering. In computational linguistics, for example, target words are represented using co-occurrences with an ordered list of context words, but semantic comparisons typically do not  depend on that order. Linguistic Matrix Theory therefore takes permutation-invariant matrix polynomials as observables \cite{LMT-2017}.

   The symmetry group governing such configurations is the finite symmetric group $S_N$, acting via the simultaneous discrete transformation:
\begin{equation}
M_{ij} \to M_{\sigma(i)\sigma(j)} \quad \text{for} \quad \sigma \in S_N
\label{eq:sn_action}
\end{equation}
Equivalently, $M\rightarrow A_\sigma M A_\sigma^T$, where $ A_{ \sigma }$ is the $N \times N$ matrix for the permutation $ \sigma $ of  $N$ basis elements of $ \mathbb{R}^N$.  The general PIGMM compatible with Eq.~\eqref{eq:sn_action} has two linear and eleven quadratic parameters and permits symmetry-controlled correlations \cite{PIGMM-2018}. It gives strong evidence for near-Gaussianity in word matrices and uses small deviations from Gaussianity in linguistic classification tasks \cite{GTMDS-2019,PIMSCLT-2022}. Permutation-invariant characteristics are likewise natural for stock and foreign-exchange correlation matrices \cite{Marti2020,PIGMFIN-2023}.

More generally, $S_N$ invariance is an exact data-reduction principle: an unrestricted Gaussian on $N^2$ entries requires $N^2$ linear and $N^2(N^2+1)/2$ quadratic coefficients, whereas PIGMM has 13 parameters, reduced to four for symmetric zero-diagonal matrices \cite{PIGMM-2018,PIGMFIN-2023}. The same symmetry perspective has recently been applied to neural-network weights \cite{PIGMAI-2025}.

\section{PIGMM: Theory and Practice  }
\label{sec:rep_theory}

For continuous $O(N)$ or $U(N)$ symmetry, diagonalisation reduces matrix integrals to eigenvalue distributions \cite{Guhr1998,Akemann2011}, which give a powerful way of studying the matrix functions invariant under the symmetries. For permutation symmetry (Eq. \eqref{eq:sn_action}),  there is no analogous symmetry reduction to eigenvalues and working directly with the rich space of invariant polynomials is the natural approach. 

\begin{figure}[t!]

  \centering
 
  \begin{tabular}{cccccccc}
    
    ~\includegraphics[scale=0.30]{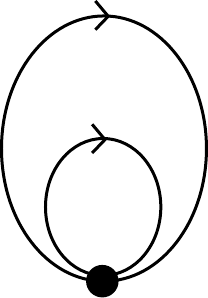}~ &
    ~\includegraphics[scale=0.30]{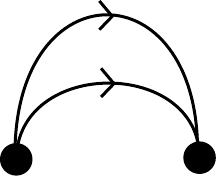}~ &
    ~\includegraphics[scale=0.30]{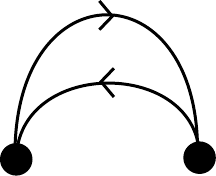}~ &
    ~\includegraphics[scale=0.30]{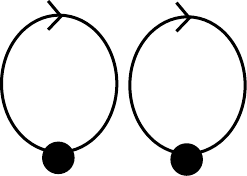}~ &    
    ~\includegraphics[scale=0.30]{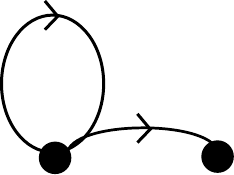}~ &    
    ~\includegraphics[scale=0.30]{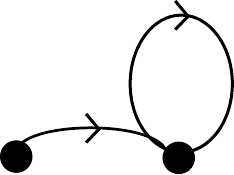}~  \\
    
    $\sum\limits_i M^2_{ii}$ & 
    $\sum\limits_{i, j} M^2_{ij}$ &
    $\sum\limits_{i, j} M_{ij}M_{ji}$ &
    $\sum\limits_{i, j} M_{ii}M_{jj}$ &
    $\sum\limits_{i,j} M_{ii}M_{ij}$ &
    $\sum\limits_{i, j} M_{ij}M_{jj}$ 
  
  \end{tabular}  

  \begin{tabular}{cccc}    
    ~\includegraphics[scale=0.30]{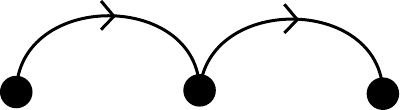}~ &
    ~\includegraphics[scale=0.30]{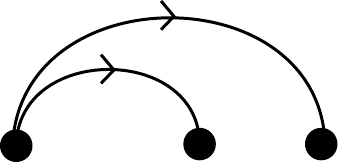}~ &
    ~\includegraphics[scale=0.30]{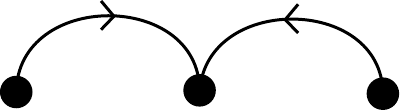}~ &
    ~\includegraphics[scale=0.30]{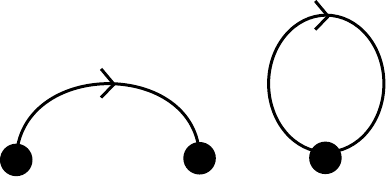} \\
    
    $\sum\limits_{i ,j , k} M_{ij}M_{jk} $ &
    $\sum\limits_{i , j , k} M_{ij}M_{ik}$ &
    $\sum\limits_{i, j , k} M_{ij}M_{kj}$ &
    $\sum\limits_{i, j , k} M_{ij}M_{kk}$
  \end{tabular}
  
  \begin{tabular}{c}
    ~\includegraphics[scale=0.30]{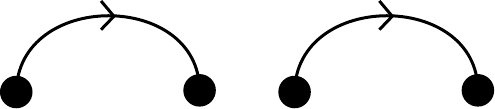} \\
    $\sum\limits_{i , j ,k , l} M_{ij}M_{kl}$
  \end{tabular}

  \caption{Eleven $S_N$ invariant quadratic functions and corresponding graphs: symmetry-controlled correlated Gaussianity  }
  \label{fig:Invts-Graphs}
\end{figure}

This viewpoint is familiar from BPS sectors of $\mathcal{N}=4$ SYM, where gauge-invariant multi-matrix operators encode thermodynamics, giant gravitons, attached open strings and back-reacted geometries \cite{CJR2001,LLM,BHLN,BBFH,KR,BHR,dMK1,dMK2,DoubCos,SR2008,SR2016}. In PIGMM, $S_N$-invariant operators correspond to directed graphs: vertices represent index labels and an edge $i\to j$ represents $M_{ij}$ \cite{LMT-2017,BPR2mat}. For large $N$, the numbers of linear, quadratic, cubic, quartic and higher observables are 
\begin{eqnarray}
2, 11 , 52 , 296 , 1724 , \cdots \nonumber 
\end{eqnarray}
The two linear invariants are $ \sum_{ i  } M_{ ii} $ and $ \sum_{ i , j } M_{ ij} $. 
The eleven graphs corresponding to quadratic observables are shown in Figure \ref{fig:Invts-Graphs}.

\subsection{Representations and the simplified action}
\label{sec:rep-action}

The graph basis gives a natural description of the permutation-invariant
polynomial functions of the matrix entries. A complementary description is
obtained by performing a group-theoretic non-abelian Fourier transform,
\begin{equation}
\left\{
\begin{array}{c}
\hbox{Graph basis for invariant functions}
\end{array}
\right\}
\quad\longleftrightarrow\quad
\left\{
\begin{array}{c}
\hbox{Representation basis for invariant functions}
\end{array}
\right\}.
\label{eq:graph-rep-Fourier}
\end{equation}
The advantage of the representation basis is that components transforming
in distinct irreducible representations of $S_N$ do not mix.

The matrix elements transform under $S_N$ according to
\begin{equation}
M_{ij}\longrightarrow M_{\sigma(i)\sigma(j)}.
\label{eq:M-SN-transformation}
\end{equation}
The transformation of a single index,
\begin{equation}
i\longrightarrow \sigma(i),
\end{equation}
defines the natural representation $V_N$ of $S_N$. This representation is
reducible and decomposes as
\begin{equation}
V_N=V_0\oplus V_H,
\label{eq:natural-rep-decomposition}
\end{equation}
where $V_0$ is the one-dimensional trivial representation and $V_H$ is the
$(N-1)$-dimensional hook representation.

Since a matrix has two indices, the matrix variables transform in
$V_N\otimes V_N$. The corresponding Clebsch--Gordan decomposition is
\begin{equation}
V_N\otimes V_N
=
2V_0\oplus 3V_H\oplus V_2\oplus V_3.
\label{eq:matrix-CG-decomposition}
\end{equation}
Here the four irreducible representations correspond to the Young diagrams
\begin{equation}
V_0=[N],
\qquad
V_H=[N-1,1],
\qquad
V_2=[N-2,2],
\qquad
V_3=[N-2,1,1],
\end{equation}
with dimensions
\begin{equation}
\dim V_0=1,
\qquad
\dim V_H=N-1,
\qquad
\dim V_2=\frac{N(N-3)}{2},
\qquad
\dim V_3=\frac{(N-1)(N-2)}{2}.
\label{eq:SN-irrep-dimensions}
\end{equation}

Representation theory therefore provides linear combinations of the matrix
elements adapted to this decomposition:
\begin{equation}
S^{V_k;\alpha_k}_{a_k}
=
\sum_{i,j}
C^{V_k;\alpha_k}_{a_k;ij}\,M_{ij},
\label{eq:rep-adapted-variables}
\end{equation}
where $V_k\in\{V_0,V_H,V_2,V_3\}$, $a_k$ labels a state in $V_k$, and $\alpha_k$ is a multiplicity index. In the stable large-$N$ regime,
\begin{equation}
\alpha_0\in\{1,2\},
\qquad
\alpha_H\in\{1,2,3\},
\qquad
\alpha_2\in\{1\},
\qquad
\alpha_3\in\{1\}.
\label{eq:multiplicity-labels}
\end{equation}
The multiplicities in Eq.~\eqref{eq:matrix-CG-decomposition} therefore produce only $2\times2$ and $3\times3$ mixing matrices. In these variables the general permutation-invariant Gaussian model has partition function
\begin{equation}
\cZ\left(
\mu_1,\mu_2;
\Lambda_{V_0},\Lambda_H,\Lambda_{V_2},\Lambda_{V_3}
\right)
=
\int dM\,e^{-\cS},
\label{eq:PIGMM-partition-function}
\end{equation}
where the action is
\begin{eqnarray}
\cS
&=&
-\sum_{\alpha=1}^{2}
\mu^{V_0}_{\alpha}S^{V_0;\alpha}
+
\frac{1}{2}
\sum_{\alpha,\beta=1}^{2}
S^{V_0;\alpha}
(\Lambda_{V_0})_{\alpha\beta}
S^{V_0;\beta}
\nonumber\\
&&
+
\frac{1}{2}
\sum_{a=1}^{N-1}
\sum_{\alpha,\beta=1}^{3}
S^{H;\alpha}_{a}
(\Lambda_H)_{\alpha\beta}
S^{H;\beta}_{a}
\nonumber\\
&&
+
\frac{1}{2}\Lambda_{V_2}
\sum_{a=1}^{N(N-3)/2}
S^{V_2}_{a}S^{V_2}_{a}
+
\frac{1}{2}\Lambda_{V_3}
\sum_{a=1}^{(N-1)(N-2)/2}
S^{V_3}_{a}S^{V_3}_{a}.
\label{eq:PIGMM-representation-action}
\end{eqnarray}
The action contains two linear and eleven quadratic parameters: symmetric $2\times2$ and $3\times3$ matrices $\Lambda_{V_0}$ and $\Lambda_H$, and the scalars $\Lambda_{V_2},\Lambda_{V_3}$ \cite{PIGMM-2018}. Its block form makes correlators straightforward to compute by Wick contractions after inverting these small matrices and scalar couplings. The algorithmic implementation of the enumeration of invariant observables, and the computation of their expectation values is available for the one and two-matrix permutation invariant Gaussians in \cite{BPR2mat}\cite{PadellaroPIG2MM}. 

\subsection{ Correlators and statistical applications} 
\label{sec:corr-stat}

Consider an ensemble $M^A_{ij}$, with $i,j\in\{1,\ldots,N\}$ and $A\in\{1,\ldots,N_{\rm ensemble}\}$, transforming under a common permutation of row and column labels. In computational linguistics, $A$ labels words represented by matrices.

For an invariant polynomial observable $O(M)$, the theoretical expectation
value is
\begin{equation}
\langle O(M)\rangle_{\rm theory}
=
\frac{1}{\cZ}
\int dM\,e^{-\cS(M)}O(M),
\end{equation}
and is compared with the empirical ensemble average
\begin{equation}
\langle O(M)\rangle_{\rm expt}
=
\frac{1}{N_{\rm ensemble}}
\sum_{A=1}^{N_{\rm ensemble}}O(M^A).
\end{equation}
Empirical averages of the two linear and eleven quadratic invariants fix the Gaussian parameters by the method of moments; Wick contractions then predict cubic and quartic observables.

A convenient measure of departure from Gaussianity is
\begin{equation}
{\rm NGM}(O)
=
\frac{
\left|
\langle O(M)\rangle_{\rm theory}
-
\langle O(M)\rangle_{\rm expt}
\right|
}{
{\rm std}_{\rm expt}(O)
}.
\label{eq:NGM}
\end{equation}
The denominator is the standard deviation of the observable as extracted from the real-world data. Applications to matrix distributional semantics found good evidence for
approximate Gaussianity in linguistic matrix data
\cite{GTMDS-2019,PIMSCLT-2022}.

For financial correlation matrices, matrix transposition symmetry and the fixed diagonal entries reduce the theory to one linear and three quadratic invariants \cite{PIGMFIN-2023}. Working with symmetric zero-diagonal matrices, the action is
\begin{eqnarray}
\cS
&=&
-\mu_{V_0}S_{{\rm phys};V_0}
+
\frac{\tau_{V_0}}{2}
S_{{\rm phys};V_0}S_{{\rm phys};V_0}
\nonumber\\
&&
+
\frac{\tau_{V_H}}{2}
\sum_{a=1}^{N-1}
S^{{\rm phys};V_H}_{a}
S^{{\rm phys};V_H}_{a}
+
\frac{\tau_{V_2}}{2}
\sum_{a=1}^{\dim V_2}
S^{{\rm phys};V_2}_{a}
S^{{\rm phys};V_2}_{a}.
\label{eq:finance-action}
\end{eqnarray}
Applied to daily financial correlation matrices, cubic and quartic invariants form a symmetry-adapted feature vector; selecting those with the largest non-Gaussianity measures gives a reduced vector. Mahalanobis distance from its mean ranks atypical days, with strong agreement with economically significant dates and performance competitive with principal-component analysis  methods \cite{PIGMFIN-2023}.

\subsection{Potential applications to particle-physics data}
\label{sec:particle-physics-applications}

The modular structure of collider simulations suggests potential applications of permutation-invariant matrix models. A typical pipeline is
\begin{equation}
\hbox{Perturbative QFT}
\longrightarrow
\hbox{parton shower}
\longrightarrow
\hbox{hadronisation}
\longrightarrow
\hbox{detector data}.
\label{eq:collider-pipeline}
\end{equation}
Correlation matrices can be constructed for in-in, out-out or in-out particle collections at each stage. Much of the physics can be expected to be insensitive to the ordering of the particles in the lists used to construct these matrices. This should provide a natural arena to study the application of the PIGMM in characterising near-Gaussianity, quantifying departures from Gaussianity, and studying the physics underlying these statistical characteristics.
The application of Deep Sets to neural network architectures in particle physics and beyond  \cite{DS1,DS2} provides an instructive precedent and a complementary perspective on this proposed use of permutation symmetry in collider physics. 

Hadron--hadron output correlations from the hadronisation step could be studied with a single-matrix PIGMM, while parton--hadron input--output correlations  naturally motivate the extension of PIGMMs to rectangular matrices with separate permutation actions; relations among several stages suggest multi-matrix ensembles. These applications could potentially lead to useful  new tools for characterising non-perturbative hadronisation phenomena and for  collider data analysis.

\begin{center}
{\bf \Large Acknowledgments}
\end{center} 
I thank the collaborators who have helped to initiate and advance the research programme on PIGMMs. I am grateful to the organizers of ACAT-2025 for the opportunity to present this talk and for the invitation to contribute to the proceedings. This work is supported by the Science and Technology Facilities Council (STFC) Consolidated Grant ST/T000686/1, ``Amplitudes, Strings and Duality''. I also acknowledge the support of a Visiting Professorship at the Dublin Institute for Advanced Studies, held during a sabbatical year in 2024, where some of the work on the research programme presented here was developed. AI tools were used to polish the final version of this contribution.

\end{document}